# User Experience, Software Interfaces, and The Unconscious

**Richard Diamond**

*Regent's College, Inner Circle, Regent's Park, London NW1 4NS, United Kingdom*
*T +44 7825 429007 E richardvladimird@gmail.com*

2 November 2009

## Abstract

Ideas about how to increase unconscious participation in interaction between 'a human' and 'a computer' are developed in this paper. Evidence of impact of the unconscious functioning is presented. The unconscious is characterised as being a responsive, contextual, and autonomous participant of human-computer interaction. Unconscious participation occurs independently of one's cognitive and educational levels and, if ignored, leads to learning inefficiencies and compulsive behaviours, illustrations of which are provided. Three practical approaches to a study of subjective user experience are outlined as follows: (a) tracing operant conditioning effects of software, (b) registering signs of brain activity psychological or information processing meaning of which is well-explored and (c) exploring submodality interfaces. Suggestions for improvement of current usability study methods, particularly eye-tracking, are offered. Conclusions consider advantages and difficulties of unconscious-embracing design and remind about evolutionary limits to a progress in built systems and environment if unconscious interaction is overlooked or obstructed.

**Author's Statement: The article is written in a format of guidance notes for a professional audience of both, practitioner and academic worlds. The article presents original work.**

## 1. Introduction: Aims

If we would like to achieve a level of seamlessness at which technology is perceived as an extension of oneself, then we need to change the square wheels of current usability management methods. In user observation, surveys and acceptance tests, we rely on such constructs as 'the ease of use' and 'perceived usefulness' which hardly changed since their early formulation in Technology Acceptance Model (Davis, 1989). At the time, they were elegant constructs that provided a well-rounded framework. After two decades, we do not have much more to offer to users and their observers in order to facilitate their awareness and improve their language in order to grasp their specific and unique experiences. We claim to design experiences but have only a few key usability constructs for a purpose of describing of human-computer interaction between. At the same time, if a technology is truly integrated with 'the unconscious' then the ease of use, perceived usefulness and usability would not be issues in themselves. Making those concepts obsolete is a criterion that an evolutionary stage is passed.

User experience of working on or with a computer is semi-conscious and organised through kinesthetics. In life, people access, verbalise and form kinesthetic experiences, but an ability to do so as a computer user is limited. Without certain practice in self-development, it is not realistic to expect to be able to access a structure of an unconscious experience, such as speech construction. In case of a difficulty, computer users refer to 'effort,' 'difficulty,' or 'confusion' in order to describe a lack of suitability of an interface for unconscious processing. What is difficult to verbalise about an experience of interaction with a computer is visible, as the user stumbles upon elements of interface design.

The issue of individual experience has another side: evolutionary development of a human kind. A connection to all parts of a human being is an imperative for any technology if it is to last and at least not to impede human functioning. If the facilitation of efficient functioning of human beings is an aim of HCI research and practice, then 'the unconscious' should be accommodated in interface design, especially as it is labelled 'experience design.' Ubiquitous computing and built environments that are not sensitive to unconscious interaction impede people rather than enabling them to be more productive or creative (Anderson & Rainie, 2008; Burleson, 2005). In order to accommodate for unconscious experience, there needs to be more specificity of how to work with unconscious processes and kinaesthetic experiences. A main aim of the paper is to provide an informative review of knowledge to that extent.

## 2.     Presence of The Unconscious: A Literature Review

Knowledge about the functioning of 'the unconscious' had been embedded into religious and healing practices of the humankind since thousands of years ago. It began to crystallise with development of the field of clinical hypnosis by Milton H. Erikson, M.D (2002). Contemporary psychological and neuroscience also learned to catch a presence of unconscious activity using both psychological experiments primarily involving memory recall (Dijksterhuis & Nordgren, 2006) and neuroscience experiments that trace brain activity that precedes conscious awareness about perception or decision (Soon et. al, 2008). In an experiment of the latter type, conducted at the Max Planck Institute for Human Cognitive and Brain Sciences, researchers predicted whether the user presses a button with their left or right hand by measuring localised brain activity (see Figure 1).

> We say that we work *on* a computer. The presupposition is similar to working on a surface. Even if we say, work *with* a computer,' the meaning would be as in working with a hammer. The conscious language overlooks interaction and particularly, unconscious level of interaction.

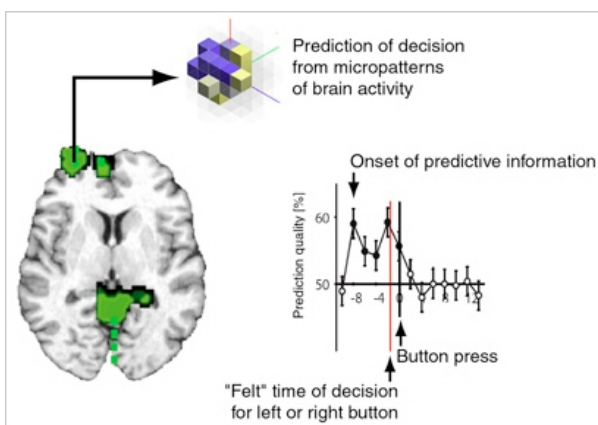

*Figure 1. Adopted from: Soon, et. al (2008)*

Schär (1996) attempted to check effectiveness of the use of conversational (auditory) interface for an explicit and conscious mode of learning and direct manipulation (kinaesthetic) interface and an implicit and unconscious mode.[1] A difficulty arose in identification whether an ill-defined tasks stimulate unconscious mode of learning. Explicit and implicit modes of learning provoke a parallel to functional specialisation between the left and right brain hemispheres.

While a proof of concept for 'the unconscous' in usability studies is not without difficulties, a concept of 'the unconscious' proves helpful in explanation of unexpected experiment results and learning inefficiencies that are repeatedly associated with IS implementation. Orlikowski (2000) repeatedly observed an early entrenchment of software use patterns in professional settings even if such patterns were sub-optimal.

Interaction between 'a human' and 'a computer' has multiple levels. In our cognitive manipulation with the interface we go in and out of awareness. There is semi-conscious intent about what we are doing or aiming to do. Visual perception is usually completely unconscious for most people, while motor functioning is unconscious to a lesser degree. As interaction with the interface goes in and out of the user's awareness, levels are created as necessary. The entire dynamics is scale-free. If one attempts to notice where is their attention at each moment, they would 'discover' quite unexpected areas of attention. Given that awareness is floating, generation of intent is semi-conscious. and the speed of motoric action is much quicker than the speed of thought, accessing and verbalising our unconscious experience represent a non-trivial problem.

The problem for designers is that any feeling of dissatisfaction with technology can remain well out of awareness. It is also common to rejecting such idea but whatever terminology is (e.g., dissatisfaction, objection, or misalignment) we can always measure delay times and muscle effort in addition to the end user's verbal reports. One might question any feeling of dissatisfaction, but such a feeling as well as 'the unconscious' can be taken as mere analytical constructs introduced for the purpose of effective explanation that allows to act. However, there is an objective verification. *Measurements of delay times, error rates, and quality of recall are also measurement of 'unconscious alignment.'* If the user successfully organised his or her unconscious resources, then they receive a quick and precise recall. Also, people use internal dialogue (auditory digital, Ad) in order to organise their consciousness and direct themselves, therefore, any internal uncertainty or argument about whether to press a button on the interface will naturally contribute to a movement delay.

Even though the construct of the autonomous unconscious makes theoretical and practical sense, in order to perceive the phenomenon of unconscious mind--to notice its signals--it might still be necessary to relax ones' objectives and adopt one or several presuppositions presented as beliefs.

## 3. Three Presuppositions of Unconscious Interaction: Framework Development

Specifically for the purposes of HCI research and interface design, I summarised the dynamics of unconscious interaction between 'a human' and 'a computer' in form of several presuppositions below. I reviewed experiment-based psychological science and particularly the complete collection of life-time work of Milton H. Erikson, MD (2002) who specialised in clinical hypnosis and exploration of unconscious functioning.

*I.   Functioning of the unconscious is autonomous. As a consequence, unconscious interaction with the computer can flow independently of one's cognitive understanding and knowledge.*

Cognitive knowledge and beliefs do affect behaviour, though their impact is better traced over time or a series of actions. All knowledge and experience--even and especially one we reject or deem unimportant consciously--are 'stored' in the unconscious mind whose autonomy is guarantee that we can draw on all richness of our experience. The protection was developed through an evolution.

Implication of this presupposition is that conscious-oriented research methods and techniques, such as surveys, have a fixed and limited potential in explication of one's full experience.

II.   *The unconscious is not bound by circumstances. Its functioning bases on all experiences, including ones that are unpleasant, forgotten and out-of-attention.*

Human behaviour is organised by contexts, such as work and home. Hypotheses were made that consciousness works from context to context and looks for a suitable context for new information. Conscious thinking applies conceptual schemes and information filtering. It is also prone to cognitive and perception constraints. Because of these fundamental characteristics of the consciousness, a therapy is needed in order to bridge across contexts and get access to psychological resources present in past or parallel experiences.

A slight change in context can significantly impact results of any usability study or psychological experiment, particularly the users ability to recall changes significantly (Altmann, 2002). A control of experimental study is largely subject to characteristics and limitations of the consciousness, including the consciousness of experimenters. An objective usability experiment cannot be performed without a tracing of the unconscious interaction.

III.  *The unconscious interacts with a computer as a living being. Unconscious responses can be characterised as literal.*

The unconscious expects responses to make sense of and react to. It can play with a computer: if a play is good there are feelings of the ease of use and enjoyment, otherwise we experience a feeling of boredom which is likely to translate into feelings of tiredness or annoyance rather quickly. A swearing at out-of-order technology reveals a presence of the unconscious interaction. Flattery from a computer produces the same general effects as flattery from humans (Johnson, et al., 2004, 2007).

It is possible to extend this list of presuppositions as well as re-combine these ideas in a generative way. It is also possible to begin to think how one can change their own design practices in order to provide for the unconscious. The next step of the paper is to consider general implications of these presuppositions for our daily work and interaction with software interfaces and computing devices.

### 3.1.   Implications. Learning Inefficiencies

The idea of the unconscious being an autonomous and engaging player is paradoxical for an information processing model of the mind which is much relied upon in usability testing, explicitly or otherwise.  Ignoring these features of the unconscious equates to ineffective behavioural learning. for at least three reasons. First, elements of contexts idiosyncratic to the user affects access to resources and skills thus, affecting productivity. Second, a badly ordered structure of stimuli confuses the unconscious which mimics sequences and cycles offered by software. Third, if the unconscious limits its responsiveness because of such confusion, the user will need to spend extra time and effort.

Learning achieved by quick impact is effective and enduring: it occurs when a certain experience threshold is passed which, in turn, associated with physiological change (Bandler & McDonald, 1986). Sudden exposure to a large concentration of hormonal stimulation is likely to create a hard-to-reverse change in neural pathways. Phobias and compulsive responses are examples such effective learning:

one does not fail to forget to produce their compulsive response. Fast learning of an undesired habit can occur regardless of whether a user consciously perceives software as useful. Software use habit(s) might represent a robust and compulsive behaviour, reversion of which requires professional help and investment--fast psychotherapy. Increasing awareness of ones unconscious experiences helps to gain alignment and integrate behaviours and internal resources into a skill; however, awareness is not necessarily sufficient to change a habit. Learning-by-doing only increases chances of unconscious integration of a skill but does not guarantee Learning II[2] in the first place.

Broken loops of software interface combined with unwillingness of the unconscious blocks learning-by-doing and creates stress and tiredness. The block to the user's ability to develop optimal sequences of behaviour by trial and error removes necessary variation in behaviour: no learning to condition and improve. Sub-optimal software use habits are common place.[3] Orlikowski (2000) reported an early entrenchment of groupware uses, of which very aim was to improve interaction. Users of office packages do not make assumptions about common functionality which would have helped them to learn about it. Users of mobile communication devices demonstrate reduced ability to navigate as if they forget screens further into a menu. A common corporate response to sub-optimal learning is to increase training expenditure and programmes. Such expenditure has uncertain return. It was only effective if end-users lacked confidence (Diatlov, 2005).

Conventional methods to study of user experience, such as demographic studies, cognitive aptitude testing, and attitudes survey do not provide specific data that would help to accommodate the unconscious. Those methods tap into surface of unconscious experience and therefore, lead to advice of a limited actionability.

## 4. Three Lines of Enquiry for Better Experience Design

Contemporary knowledge about unconscious functioning allows to identify three following approaches to an explication of a structure of user experience:

- Analyse *software as a trainer* that modifies behavior via operant conditioning.

- Trace *observable* signs of a functional specialisation of brain areas.

- Explicate submodatlities--interfaces that the unconscious uses to structure thinking.

### 4.1. Tracing Operant Conditioning Effects

It is easy to see how software is rich in sequences and loops of stimuli. If a user does not follow them, they are punished with an error, crash, loss of data, and extra work. Software modifies our behaviour. A mechanism through which it does it is called operant conditioning. It is based on sequences of rewarding or punishing stimuli, called schedules (Skinner, 1968). Self-reinforcement plays a sizeable role operant conditioning: the unconscious begins to anticipate repeating stimuli, adopt behaviour in order to get more of them or avoid them, and in some cases, mimic a stimulus in order to get a desired response from itself. The last characteristic is problematic: mimicking a stimulus works on humans but not on computers.

Counter-productive consequences of a misuse of operant conditioning are observed in contemporary use habits. First and frequent consequence is *compulsion*. It is easy to find users who consciously know that software is configured to check their e-mail box automatically, hear an auditory notification of a new mail regularly, still click on 'Receive' button compulsively in order 'to stimulate' software to get more e-mail. That is also an example of 'magical thinking': as if pressing a button will generate new email by itself. Conscious knowledge does not prevent an unconscious response.

Second counter-productive consequence is *amnesia* that occurs as a result of violation of established sequences of behaviour. For example, if a software wizard ends without a review of options then there is less chance that the user would remember them. A review of options provides a pause and chance of behavioural control preventing a feeling of being lost reported frequently otherwise, particularly in cases of software configuration. While outcomes of amnesia are not always negative, it makes a reflexive learning difficult and requiring extra time and effort. Altmann's (2002) near-term memory model confirmed that cluttered interface might impose cognitive costs by increasing retrieval demands on memory.

Operant conditioning is a powerful phenomenon that should be accommodated. Behaviours entrained via operant conditioning remain even as the users' needs and circumstances change. Two scenarios described in this section are common. Therefore, it is important to check an UI for operant conditioning effects. To help that, it is possible to create a software that traces interface events and identifies schedules of operational conditioning. Such software would help to trace development of use habits and skills.

## 4.2. Registering Signs of Known Brain Activity

Exploration of linkages between brain activity and specific conscious functions is a broad and contested area. Even so, such linkages are commonly implied in usability research.

Results of brain scanning are open to multiple interpretations because a variety of technologies and methods is used[4] and impact of experimental conditions is difficult to trace and extrapolate on out of the lab contexts. Brian activity offers multiple dimensions and ways in which it can be separated and matched with a specific cognitive activity.

**Eye Tracking**

Eye tracking is commonly applied in usability testing, in which eye fixation (gaze) and its duration are traced in order to trace 'attention.' Research questions are asked as follows: Did the user overlook element X? Were they distracted by another element Y. Answers to such questions tell little about subjective user experience, ease of use, and satisfaction--a conventional usability framework (Davis, 1989). Resulting data is volumous and practitioners recognsie its low usefulness (Namahn, 2001).

Linking eye fixation and conscious activity of 'attention' is a tenuous presupposition, even if experiment subjects are instructed focus on the interface. It would be valid if 'the unconscious' did not mediate that linkage, having its own purposes, such as gathering information about other crucial elements of the environment. Looking for and measuring of conscious attention is an approach that stems from conscious thinking itself. Another tenuous presupposition would be that conscious focus is important for making a decision. As illustrated above, that is not usually the case: intense unconscious activity is registered before a felt time (conscious realisation) of a decision (Figure 1).

Direct gaze is unnecessary for the unconscious to notice an element and adjust one's reaction accordingly. The unconscious grasps many elements at once and reacts to them selectively using its own criteria of importance. The unconscious have no limitations as to the limited conscious information processing capacity of 7±2 chunks of information. It is able to observe and process a bigger picture. It is also uses such perception distinctions as between a figure and background in order to sort thoughts.

The autonomy of unconscious functioning is a key presupposition for the interpretation of eye-tracking: eye movement is not only a response to external stimuli but also reflects internal brain activity. Particularly, eye movement is revealing of 'a dialogue' between the left and right hemispheres.

**Left and Right Brain Hemispheres**

Our current knowledge in this area was trivialised: stereotypes about 'left brain' and 'right brain' were formed and proven wrong. That does not mean that there are no differences in how brain hemispheres process information or there is no domination--at any given time, one hemisphere sends more intense signals than the other.[5] The dynamics between the left and right brain hemispheres can provide a unique and reliable base for insight into unconscious functioning and experience. First, precisely because neurological research focused on the area, there is a better knowledge. Second, that left-right hemispheric dialogue is architecturally central to the brain and higher order cognitive functioning. Third, a left-right distribution of hand, head, eyes and even mouth movement can be traced by a human observer. A dialogue--or as some research put it *domination*--between the left and right brain hemispheres is a kind of brain activity that can be observed directly and reliably.

**Expression Tracing Technologies**

It is possible to observe the left-right expression, particularly eye and head movement and match it with speech content. Such tracing would aim to identify how a subject sorts phrases and individual words to the left or right, and then, identify categories of content that are predominantly sorted to one side. That can be done using frequency statistics and will provide a reliable indication of brain hemispheric activity. Observational data exist that certain categories of meaning 'belong' to a certain hemisphere, for example, right-handed people sort past to the left and future to the right. Such data, being statistically verified, will give a new ground for an interpretation of user experience. It will be possible to make a practical assessment of whether an interface is conductive to a purpose, if we know a kind of participation of a particular hemisphere in a fulfillment of that purpose and whether information processing features of that hemisphere are suitable with the task and sequence of interface elements.

Current eye-tracking analysis already tries to fuse such data as mouse movement, facial expression, and voice input. The accessing cues model developed in the area of neuro-linguistic programming can provide an initial mapping to what is processed in which hemisphere. The model suggests other, less known facts, that can help a practical tracing of expression, or example, mouth and head movement help eye movement.

In the future, expression tracing can be improved by 3D body imaging, which operates with a cloud of points that measures human body (Treleaven et. al, 2006). Such scanning can be performed at a necessary frequency in order to catch up with a change in expression, but necessary algorithms, code and data storage for such a task are not there yet. Current commercial developments, such as robot toys achieve impressive results in human expression tracing--and responding with a simulated artificial intelligence--through a combination of simple motion detectors.

### 4.3. Exploring Submodality Interfaces

Elements of interface provide metaphors that structure one's experience, for example, expression 'went blank' describes a state of a human as well of a computer. Such metaphors can be literal and powerful instructions that are taken on an unconscious level of perception. Perhaps, those features of computer interfaces and architecture are projections of the unconscious functioning in the first place. It is a registered phenomenon that the unconscious can assign (arbitrary) meaning to visual, auditory and kinaesthetic effects and then utilise those effects in order access meanings and related experiences. The contemporary software interfaces provide a rich selection of such effects to the unconscious.

> IV. *'The unconscious' utilises invariants of perception in order to sort subjective experiences. The invariants are submodalities that characterise each representational system--that are Visual, Auditory, Kinaesthetic, Olfactory and Gustatory (VAKOG).*[6]

Change in visual (V) submodalities of colour, size and proximity creates a change in our experience. A close, large and panoramic picture creates a stronger kinesthetic response than a small and distant one. The effect is subject to a threshold: if a picture is too large or too small it might have no effect because it is not noticed. A distinction between background and a picture defines is another invariant that defines such a threshold. In advertising and visual communications, manipulation with size and proximity of a picture is a common and reliable way to convey significance and stimulate an unconscious response. Utilisation of submodalities can be presented as 'access' to feelings but is, in fact, a complex process process that might involve associations of meaning.

Another common element of visual communication and computer interfaces is a morphing of one interface element into another. A morphing signals association. If such an association is novel to 'the unconscious' it takes time and effort to process it: practically impossible associations are not just disregarded by the unconscious--as 'a conscious mind' would do. Excessive use of morphing and disarray of interface elements create amnesia. It is not infrequent to observe a user who is lost on a familiar window or, worse, almost always experiences amnesia when seeing a certain window. That phenomenon leads to slowing down and increased effort of the user.

Manipulation with submodalities can be astoundingly effective. 3D visual in combination with a receipt of correlated multisensory information is sufficient to trigger an illusion that another person's body or an artificial body was one's own (Petkova & Ehrsson, 2008).

Submodalities themselves develop as a result of exposure to technological interfaces. Without exposure to cinema screens we might not have had such a strong response when seeing a large and panoramic picture. Technology allows for a creation of new submodalities and a better manipulation with submodalities that are already utilised by the unconscious. Submodality creation occurs as the unconscious seeks for invariance in VAKOG context and utilises those patterns in order to assign and keep track of a particular meaning of significance. For example, we all know a distinction between a blue day and bright day.

A purposeful manipulation with submodalities facilitates access to unconscious processes as needed. It is thus possible to attune software interfaces in order to focus attention, facilitate entrainment of productive habits and activate necessary strategies of information processing--all with a minimal conscious intervention. An argument could be made that technology is getting control over 'the user.' It is, however, known that effective, effortless and enjoyable skill exercise occurs on its own, with a minimal conscious participation. Anyone who had and experience of a sports activity or gardening can find a confirmation of that. After all, we formulate sentences of language and pronounce them with a minimal conscious participation as well.

## 5. The Unconscious In HCI Research: A Deductive Interpretation Example

In order to provide a further illustration of how these ideas about unconscious functioning can help HCI research, a study by Tan, et. al (2002) is interpreted below. The study found 19% improvement in spatial memory for information accessed via a touchscreen versus mouse. In addition, it was found that females are likely to benefit more than males from using a touchscreen device.

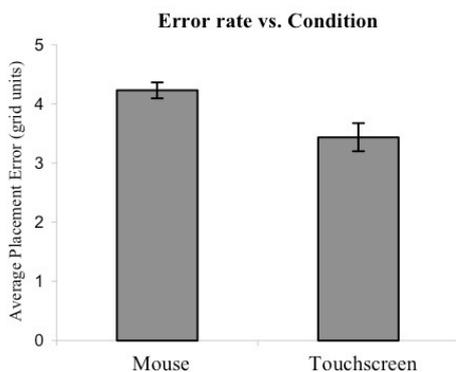
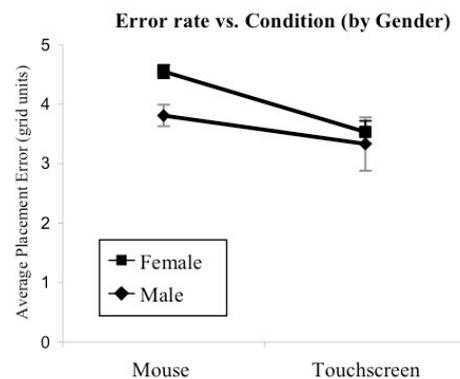

**Figure 1:** Touchscreen users performed significantly better on the spatial recall test.

**Figure 2:** Females benefited more from using the touchscreen than it did males.

*Figure 2. Adopted from: Tan et al. (2002)*

Gender differences in experiment results can be attributed to a better development of or access to the right-hemispheric information processing by females. The right hemisphere is associated with kinaesthetic representations. It is also possible to say that this aspect of hemispheric specialisation contributed to the reduction of error rate by females.

Use of a touchscreen stimulates and simplifies the unconscious functioning by offering a possibility of mapping across the visual and kinaesthetic representation systems. Stimulation: direct manipulation interfaces induce an implicit (unconscious) mode of learning (Schär, 1996). Simplicity: V-K synaesthesia facilitates orientation on the outside as well as access to memory on the inside, simply because more neural and brain circuitry is involved in the process. It can also be hypothesised that a redundancy of simultaneous processing in two representational systems leads to a more integrated dialogue of the left and right hemispheres, which makes it easier for the unconscious to generate a solution (a right movement). V-K synesthesia of handwriting is one of the most trained in schools. Effect of an improved recall was also facilitated by direct spatial anchoring--that is, unconscious use of location submodality in order to sort various categories of significance.

Use of a mouse occurs on a surface that is perpendicular to a screen. The indirectness leads to decreased V-K synaesthesia and increased the amount of brain computational resources needed to co-ordinate mouse movement. Given that the visual and kinaesthetic representation systems operate in a more separate way, navigation with a mouse is prone to interruption end error. It is a common phenomenon to forget about a pointer's position. Thus, a quick and small forward and backward movement of a mouse becomes an additional but necessary habit. Additional co-ordination load on the unconscious decreases its memorising and recall abilities. The co-ordination between the left and right is genetically determined and its breach leads to negative kinaesthetics and unconscious, hard-to-explain blockages in memory, motivation and motor responses. However, we manipulate with objects on the left using the right hand on Microsoft Windows desktop, where Start menu and icons are aligned to the left by default and most users are right-handed.

## 6. Conclusions

The paper provided advice on creation of interfaces to which our unconscious minds would want to connect. The unconscious mind is an autonomous, contextual, responsive and sometimes playful independent thinker. The paper presented these basic presuppositions of the unconscious functioning and illustrated how appreciation of them can and update our prospective on human-computer interaction. Three lines of enquiry for better experience design were offered together with potential techniques and software tools.

*An argument might be made that existing usability research accounts for unconscious effects or intentionally choose not to hypothesise mental processes.* If the unconscious is an independent thinker then only outcomes of those thinking are registered and could easily be misinterpreted without a conceptual map that accounts for the unconscious functioning, which is too powerful, physiology-regulating agent to ignore. Evolutionary protection of the autonomy of the unconscious functioning also contributes to a tacit nature of its impact. Unconscious objections appear in tacit ways and over time. A typical objection occurs as a loss of interest that might not be apparent to the user themselves. Confusion, perception of inconvenience and delays are also signs of a lack of unconscious acceptance. Users do not feel a need to rationalise such objectives; nether they are able to talk about their sources. For a purpose of measurement, response delay is a more reliable signal of unconscious objections rather then error rate. The latter could be a function of training.

*Another argument might concern costs and complexity of an unconscious-embracing design.* Current usability testing and research are expensive even without a comparison. It is also difficult to evaluate a return on investment from common methods, such as ethnography and surveys. Ethnographic studies of the use of hardware and software by various social groups and in various contexts of life are lengthy, expensive and generate rich and unstructured data, for which a basis of interpretation are lacking. Only big organsiations, such as Intel Corporation, can employ a team of in-house PhD-level ethnographers as their 'secret weapon.'[7] Productivity and return on eye-tracking are explicitly doubted by industry experts. Results of eye-tracking also remain under-utilised without a good interpretation basis. Lines of enquiry into unconscious experience presented in the article can also generate data that is volumous and idiosyncratic to individual preferences and needs. However, such data can supplement and improve interpretation of existent methods and research results, therefore improving a quality of modelling output. Acknowledging presence of 'the unconscious' could be a relatively minor but very useful adjustment. The presuppositions of the unconscious functioning can be used as guiding principles in any design of interface or user experience study.

*Perhaps an overarching answer to a discussion of unconscious-embracing design is a proposition that human evolutionary choices are involved.* Most of existent software architecture, programming and design are products of the left-hemisphere thinking stimulating left-hemisphere kinds of information processing in users.[8] A direct verification of this claim is an open issue for further empirical research. However, a one-sideness of human experience is phenomenologically observed as today's ever-capable computing technology is associated with a decrease in creativity, uniqueness of generated content and, as a result, a loss of certain business, cultural and other kinds of value. To what degree, contemporary interfaces themselves stimulate such a negative effect is another issue for further research.

One-sided experience refuses information to that magnificent learning machine that a human brain is. Overlooking the unconscious removes unconscious motivation which is necessary for learning and self-actualisation (Burleson, 2005). It is not a coincidence that a review of knowledge on the unconscious interaction between 'a human' and 'a computer' includes learning inefficiencies and compulsive behaviours. The extend of such inefficiency signifies a potential gain that can be made if 'the unconscious' is accommodated for. The potential should be used in order to enhance our behavioural competencies rather than impede them. We can use submodalities, expression tracing and operand conditioning in order to stimulate types of learning and information processing that we need. The first task of unconscious-embracing interface design is to remove compulsive cyclic behaviours. It is necessary to clean up our interaction with technology and built environment. Improving suitability of interfaces to our unconscious processing will lead to faster, seamless and enjoyable work. Accommodating for the unconscious is more than an issue of our habits and likes. It is an issue of building an environment that supports human evolution and takes into account our ancestry. User experience and software interfaces certainly look different from this perspective.

## Notes

[1] Explicit and conscious mode is characterised by a selective attention on a problem. Implicit and unconscious mode is based on trial and error (Schar, 1996).

[2] Learning II is learning how to learn.

[3] Sub-optimal learning is one that stops/entrenches before the best best way is learned or aim of the learning is achieved.

[4] Such names as Magnetic Resonance Imaging (MRI) serve as umbrellas for a variety of technologically diverse methods.

[5] Extreme example of a shutdown of one hemisphere and bursty another is a phobic response. Co-incidentally phobias are an example of instant and endurable learning.

[6] For purposes of a study of a structure of subjective experience, any context can be described in terms of VAKOG. It can be external or internal. Internal context is formed by states of mind and nonprioreceiptive kinethetics.

[7] For examples of ethnographic work at Intel Corporation please see http://www.intel.com/research/exploratory/papr/

[8] Whether 'left-hemisphere thinking' is factual or model term is of less relevance. The term is used to identify a phenomenological set of specific informaiton-processing strategies that lean towards formal logics, linear connections and deterministic causality.